\documentclass[reprint,amsmath,amssymb,aps]{revtex4-1}

\usepackage{graphicx}% Include figure files
\usepackage{dcolumn}% Align table columns on decimal point
\usepackage{bm}% bold math
\usepackage{color}

\begin{document}

\preprint{APS/123-QED}

\title{Extremely low inhomogeneous broadening of exciton lines\\
in shallow (In,Ga)As/GaAs quantum wells}

\author{S.V. Poltavtsev}
 	\email{to whom the correspondence should be addressed: svp@bk.ru}
	\affiliation{Spin Optics Laboratory, St.-Petersburg State University, 1 Ul'anovskaya, Peterhof, St.-Petersburg 198504, Russia}
 
\author{Yu.P. Efimov}
\author{Yu.K. Dolgikh}
\author{S.A. Eliseev}
\author{V.V. Petrov}
\author{V.V. Ovsyankin}
 	\affiliation{Deprtment of Physics, St.-Petersburg State University, 1 Ul'anovskaya, Peterhof, St.-Petersburg 198504, Russia.}

\date{\today}

\begin{abstract}
We study radiative linewidth of exciton resonance in shallow In$_x$Ga$_{1-x}$As/GaAs single quantum wells as a function of indium concentration in the range $x=0.02...0.10$ and well thickness in the range $L_Z=1...30$ nm using the method of Brewster reflection spectroscopy. Record linewidths of heavy-hole exciton resonance of about 130...180 $\mu$eV are measured in reflection spectra for single quantum wells with $L_Z=2$ nm and $x=0.02$ at temperature 9 K. In these spectra, the non-radiative linewidth including inhomogeneous broadening can be comparable or even less than radiative linewidth. It is shown that radiative linewidth weakly depends on $x$ and $L_Z$ in these ranges. In multiple-quantum-well Bragg structure with ten periods radiative linewidth exceeds inhomogeneous broadening by 4 times. 

\begin{description}
\item[PACS numbers] 42.50.Ex, 42.62.Fi, 42.25.Gy, 42.70.Nq
\end{description}

\end{abstract}


%\keywords{MQW Bragg structures, Brewster angle, reflectance, inhomogeneous broadening}%Use showkeys class option if keyword display desired
\maketitle

\section{\label{sec:level1}Introduction}

Semiconductor quantum-well-based devices play an important role in optical information transfer and laser technologies. The nano-structures with III-V quantum wells (QW) are considered as promising candidates to basic medium of future elements of information processing in a pure optical way \cite{Oves1,Oves2}. In spite of the fact that molecular beam epitaxy (MBE) is highly developed, however, the quality of heterostructure growth achieved nowadays is insufficient for some perspective applications.

In the context of optical information processing, the most important properties of the QW system are the rate of resonant interaction with light and efficiency of coherence transfer from resonant light to quasi-2D excitons and back, from excitons to light. While the interaction rate is promisingly high due to the large oscillator strength of allowed exciton transitions in QWs as compared to other resonant systems (alkali atoms, impurity centers etc.), the coherence transfer is typically inefficient because of the dominant role of non-radiative damping processes present in QW, which are stronger even at low temperatures. The coherence transfer efficiency $F$ can be expressed by the ratio of teh radiative decay rate of excitons $\Gamma_0$ to the total decay rate of exciton ensemble including radiative and all non-radiative mechanisms ($\Gamma_{NR}$), such as exciton-phonon interaction, scattering on potential disorder and impurities, spread of individual oscillator frequencies etc., so, $F=\Gamma_0/(\Gamma_0+ \Gamma_{NR})$. For the potential optical processing applications $F$ should be close to unity meaning dominant role of $\Gamma_0$.

To achieve this goal, two basic approaches seem feasible. The first one is reduction of role of exciton non-radiative damping mechanisms. While temperature-dependent exciton-phonon scattering can be sufficiently suppressed by cooling down the sample, the major non-radiative mechanism remains: static potential disorder is defined by the growth procedure and results in inhomogeneous broadening of the exciton absorption line, which in III-V compounds typically exceeds radiative linewidth by several times. The reduction of inhomogeneity factor is a challenging task of MBE technology, and in this paper we demonstrate, in particular, that further technological efforts are reasonable.

The second approach is directed to the enhancement of oscillator strength, which is linearly related to $\Gamma_0$. For single quantum well (SQW), exciton oscillator strength is a function of two main parameters -- thickness of the well layer $L_Z$ and concentration $x$ in Al$_x$Ga$_{1-x}$As and In$_x$Ga$_{1-x}$As alloys. For the (Al,Ga)As/GaAs SQW structures, where heavy-hole exciton wave-function is highly confined within the well layer, the exciton oscillator strength is higher for smaller well thicknesses \cite{Voliotis, Poltavtsev1, Pasquarello}, although inhomogeneity for narrower quantum wells is higher too. Excitons in (In,Ga)As/GaAs SQW reveal tendency to a reduction of oscillator strength with decreasing well thickness \cite{DAndrea, Zhang}, because the exciton wave-function easily penetrates the barriers and confinement-enhanced Coulomb interaction between electron and hole reduces. For both systems, the reported linewidths of heavy-hole exciton resonance are always a few times broader as compared to the radiative linewidth \cite{Zhang, Prineas1}, which is of the order of 60 $\mu$eV \cite{Prineas2}.

Two decades ago, Ivchenko and coauthors \cite{Ivchenko1} suggested a conception of multiple-quantum-well (MQW) Bragg structures with the quantum well layers separated from each other by the half-wavelength period. It has been theoretically shown that for the geometry of normal incidence of light, the oscillator strengths of individual quantum wells are accumulated and radiative part of reflection spectrum gets wider in proportion with the number of QWs, $N$. Later on, it was experimentally demonstrated by the reflectance spectroscopy that in In$_{0.04}$Ga$_{0.96}$As MQW Bragg structures the total linewidth, indeed, grows linearly up to $N=100$, and the resonant reflection coefficient increases up to 95\%, for samples with anti-reflection coating and normal geometry of measurement \cite{Prineas2}. Sample with $N=210$ demonstrates close to 100\% reflection \cite{Prineas1}. Time-resolved measurements by the degenerate four-wave-mixing revealed presence of fast decaying super-radiant component from high-quality $N=10$ MQW Bragg structure confirming efficient radiative coupling of excitons of different QWs \cite{Hubner1}. While the radiative linewidth increases linearly with $N$, the non-radiative broadening is associated with fluctuating nature of local inhomogeneity and increases sub-linearly, hence, the latter affects mainly on the reflection spectra of low-$N$ structures. The influence of disorder on absorption spectrum of MQW Bragg structure is carefully studied in theory \cite{Kosobukin}. Thus, it is technologically feasible to manufacture QW-based structure with fast operating rate and the coherence transfer efficiency $F\approx1$, but in a rather complex way.

In this paper, we report on results obtained with reflection spectroscopy performed at $T=9$ K on a set of (In,Ga)As/GaAs SQW structures demonstrating extremely low inhomogeneous broadening of excitonic lines comparable with radiative linewidth. We also study $N=10$ MQW Bragg structure, where radiative linewidth four times larger than the total broadening of exciton line. Background-free reflection spectra measured in Brewster geometry exhibit exciton linewidths as narrow as 130...180 $\mu$eV for SQW samples and resonant reflection coefficient as high as 0.7 for MQW sample.

\begin{figure}[t]
\includegraphics[width=\linewidth]{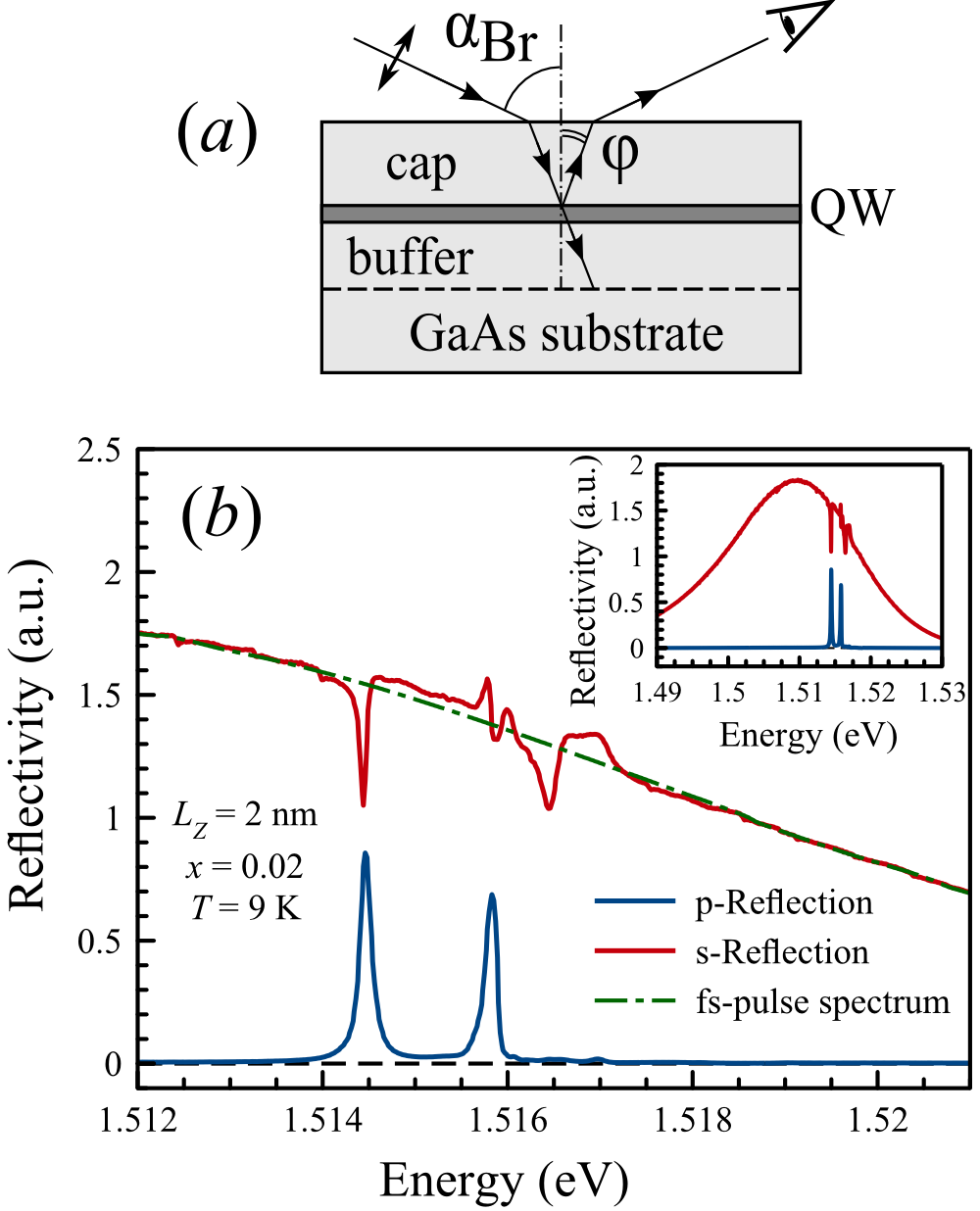}
\caption{(Color online) The Brewster reflection spectroscopy method: (a) geometry of the experiment; (b) The spectral data for 2 nm In$_{0.02}$Ga$_{0.98}$As/GaAs SQW (P561) explaining measurement procedure: {\it blue and red solid} lines are raw reflection intensity spectra obtained in {\it p}- and {\it s}-polarizations; {\it green dash-dotted} line is smoothed fs-pulse spectrum used for correction resultant excitonic spectrum. The inset shows the spectra in a wide spectral range displaying, in particular, the 20 meV-wide spectrum of fs-pulses.}
\label{method}
\end{figure}

The paper is organized as follows. In {\bf section II}, experimental method used for background-free reflection spectra measurements is explained, and a set of samples under investigation are described. Then, theoretical model used for spectral data simulation is given. Results of the measurements performed on SQW and MQW samples are presented in {\bf section III}. Conclusion is given in {\bf section IV}.

\section{\label{sec:level1}Experimental method and theoretical model}

To measure reflectivity spectra from the QW samples the Brewster angle geometry and femtosecond laser pulses have been exploited. The samples were mounted in a closed-cycle cryostat and cooled down to 9 K. The linearly polarized laser beam was focused to a spot of about 50$\times$150 $\mu$m at the Brewster angle of incidence, $\alpha_{Br}=74.5^{\circ}\pm0.5^{\circ}$, as it is shown schematically in Fig.~\ref{method}(a). The reflected beam was focused to the open slit ($\approx$0.5 mm) of a 0.6-meter spectrometer with grating 1200 lines/mm, equipped with CCD at the output. Spectral resolution provided by the set-up is 20 $\mu$eV. The wide spectrum of 100-fs pulses of the Ti:sapphire laser (Tsunami) allowed one to cover all spectral features of interest in the sample. Intensity of fs-pulses used in the experiment was checked to lie in the range of linear excitation regime. To get correct reflectivity spectrum, the data measured in {\it p}-polarization were divided by the spectrum of fs-pulses measured in s-polarization, which has been smoothed to eliminate all spectral features of the QW as shown in Fig.~\ref{method}(b) (dash-dotted line), taking into account the Fresnel reflectivity in {\it s}-polarization from the GaAs material and the value of spectrometer dichroism. The residual coefficient of the off-resonant reflection, typically of about 0.0012, is due to slight mismatch between the refraction indices of QW and barrier layers, and finite aperture of the focused beam. Previously, it was shown that the time-integrated reflection spectra measured with spectrally wide fs-pulses are similar to those measured with a tunable cw-laser source \cite{Poltavtsev2}. 

The samples under study were 20 (In,Ga)As/GaAs heterostructures comprising total 36 SQWs and one N~=~10 MQW Bragg structure grown on (100) GaAs substrates (see Table \ref{tabular:samples_data} for details). The growth conditions and sample structure were optimized during the manufacturing and characterization processes, which resulted in the following simple structure exploited for all high-quality samples described in the paper: GaAs substrate was overgrown by GaAs buffer layer of ~1 $\mu$m thickness followed by one to five In$_x$Ga$_{1-x}$As QW layers having $L_Z=1\div31$ nm separated by GaAs barriers of $50\div110$ nm thickness and terminated by $17\div220$ nm GaAs cap layer. Indium concentration in the QW layers was ranging between 0.02 and 0.10. The sample of MQW structure (P555) was composed of ten 3 nm-thick In$_{0.02}$Ga$_{0.98}$As QWs separated by 110 nm-thick barriers nominally satisfying Bragg conditions calculated for the Brewster angle geometry. The samples were grown without substrate rotation and possess a high spatial inhomogeneity of spectral properties (with typical spatial scale exceeding 200 $\mu$m)
caused by a variation of layer thicknesses and indium concentration.

\begin{table}[h]
\begin{center}
\footnotesize
\begin{tabular} {|l|c|c|l|c|c|} \hline
\bf sample, QW & \bf $\boldsymbol L_Z$ (nm) & $\boldsymbol x$ & \bf sample, QW & \bf $\boldsymbol L_Z$ (nm) & \bf $\boldsymbol x$ \\ \hline
P531, QW1 & 2.5 & 0.05 & P549, QW2 & 3.2 & 0.03 \\ \hline
P531, QW2 & 3.5 & 0.05 & P549, QW3 & 2.0 & 0.03 \\ \hline
P531, QW3 & 7.0 & 0.05 & P549, QW4 & 1.2 & 0.03 \\ \hline
P531, QW4 & 9.5 & 0.05 & P550, QW1 & 2.0 & 0.03 \\ \hline
P531, QW5 & 31.0 & 0.05 & P550, QW2 & 3.3 & 0.03 \\ \hline
P532, QW1 & 2.0 & 0.05 & P550, QW3 & 4.0 & 0.036 \\ \hline
P532, QW2 & 4.0 & 0.05 & P551 & 3.0 & 0.035 \\ \hline
P536 & 1.0 & 0.09 & P552 & 2.0 & 0.02 \\ \hline
P538 & 2.0 & 0.05 & P553 & 2.0 & 0.02 \\ \hline
P540, QW1 & 4.0 & 0.08 & P554 & 2.0 & 0.02 \\ \hline
P540, QW2 & 2.0 & 0.08 & P557 & 20.0 & 0.02 \\ \hline
P544 & 2.0 & 0.10 & P561 & 2.0 & 0.02 \\ \hline
P546 & 4.1 & 0.05 & P564 & 4.0 & 0.03 \\ \hline
P547, QW1 & 2.1 & 0.04 & P565 & 4.4 & 0.03 \\ \hline
P547, QW2 & 4.4 & 0.04 & P566, QW1 & 2.0 & 0.02 \\ \hline
P548, QW1 & 1.8 & 0.04 & P566, QW2 & 3.0 & 0.02 \\ \hline
P548, QW2 & 1.8 & 0.07 & P566, QW3 & 4.0 & 0.02 \\ \hline
P548, QW3 & 1.8 & 0.10 &  &  &  \\ \hline
P549, QW1 & 4.2 & 0.03 & P555 (MQW) & 10 x 3.0 & 0.02 \\ \hline
\end{tabular}
\end{center}
\caption{\label{tabular:samples_data} Growth parameters of the samples under study.}
\end{table}

To simulate experimental reflection spectra we used the transfer matrix method based on several approximations. First, we neglect the difference of background permittivity $\varepsilon_0$ of different layers, since we see no reflection far from any resonances. Second, we exploit Lorentz approximation implying that all mechanisms of exciton ensemble relaxation can be characterized by certain relaxation rates, including spread of individual oscillator frequencies, which leads to inhomogeneous broadening of reflection spectrum. In case of significant inhomogeneity, this approximation would be insufficient and analysis based on non-local dielectric response, given for instance in \cite{AKavokin1,AKavokin2}, would be required. Here in our study, we consider this inhomogeneity to be insufficient to distort Lorentz line shape of absorption. To describe resonant features we used following local effective single-pole dielectric function \cite{Ivchenko2, Ivchenko3}:

\begin{equation}
 \begin{split}
 \varepsilon(\omega)=\varepsilon_0+\varepsilon_0\omega_{eff}&/\left(\omega_0-\omega-i\Gamma_{NR}\right),\\
 \omega_{eff}=2\Gamma_0/&sin\left(\omega\sqrt{\varepsilon_0}L_Z/c\right).
\end{split}
\end{equation}

\noindent Here, $\omega_0$ is the heavy-hole exciton resonance frequency, $\omega$ is the light frequency, $\Gamma_0$ and $\Gamma_{NR}$ are the radiative and non-radiative decay rates of excitons, respectively, $\omega_{eff}$ is the effective longitudinal-transverse splitting of excitons, and $c$ is the speed of light. In particular cases, when the exciton line is an isolated peak in background-free spectrum [similar to Fig.~\ref{method}(b)], its shape is typically very well fitted with Lorentz line allowing one to conduct a simple analysis as it has been done previously on GaAs SQW \cite{Poltavtsev1}. The resonance shape can be described then with the following simple formula \cite{Ivchenko1}:

\begin{equation}
R(\omega)=\frac{\Gamma_0^2}{(\omega-\omega_0)^2+(\Gamma_0+\Gamma_{NR})^2}.
\end{equation}

Using the resonant reflection coefficient $K_{RR}=R(\omega_0)$ and the value of full-width at half-maximum (FWHM) of the exciton line $\Gamma$, we can obtain radiative and non-radiative exciton linewidths, which we associate with according exciton decay rates:

\begin{equation}
K_{RR}=\frac{\Gamma_0^2}{\Gamma^2},
\end{equation}

\begin{equation}\label{KRR}
\begin{split}
\Gamma=\Gamma_0&+\Gamma_{NR},\\
K_{RR}=1/(1&+\Gamma_{NR}/\Gamma_0)^2
\end{split}
\end{equation}

\begin{equation}\label{Gamma_zero}
\begin{split}
\Gamma_0=&\Gamma\sqrt{K_{RR}},\\
\Gamma_{NR}=\Gamma&(1-\sqrt{K_{RR}}).
\end{split}
\end{equation}

It should be noted that the Brewster angle geometry used in our experiments differs from the normal incidence geometry implied by this simple theoretical approach, but although the angle of light incidence on the sample surface is large ($\approx74.5^{\circ}$), the angle inside the structure is fairly small [$\varphi\approx16^{\circ}$, see Fig.~\ref{method}(a)] due to high background refraction index in GaAs layers, $n_0\approx3.5$. According to Andreani et. al. \cite{Andreani}, the radiative contribution to the exciton reflectance will be slightly decreased proportionally to $cos \varphi$, which in our case is within 4\% of accuracy and we neglect it for simplicity.

\section{\label{sec:level1}Experimental results}

\subsection{\label{sec:level1}Resonant reflection from SQW}

\begin{figure}[t]
\includegraphics[width=\linewidth]{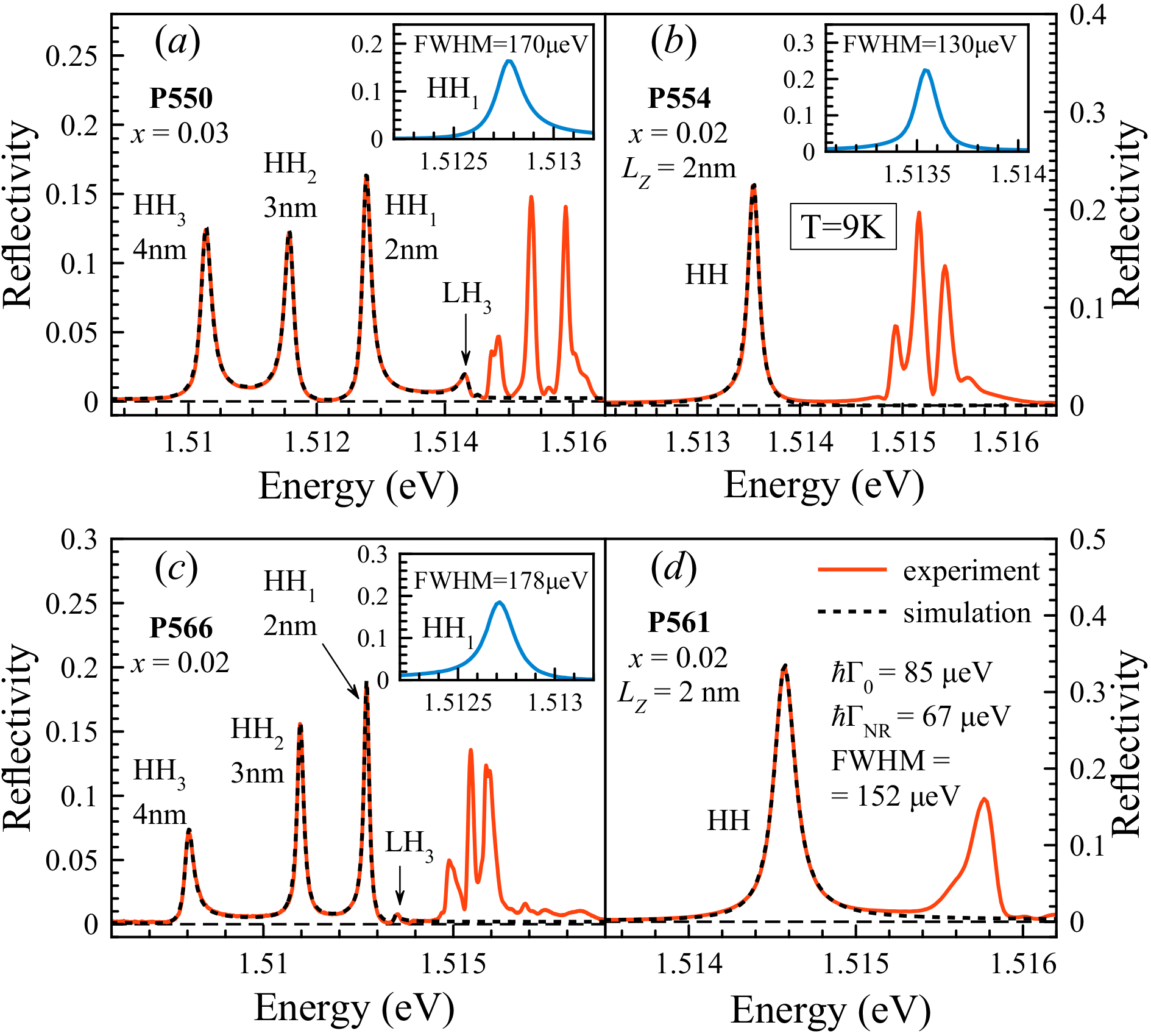}
\caption{(Color online) Reflectivity spectra of the four SQW samples measured in Brewster geometry ({\it red solid} - experiment; {\it black dash} - simulation.): (a) sample P550: $\hbar\Gamma$(QW1)=170 $\mu$eV, $\hbar\Gamma$(QW2)=181 $\mu$eV, $\hbar\Gamma$(QW3) = 211 $\mu$eV; (b) sample P554: $\hbar\Gamma$=130 $\mu$eV; (c) sample P566: $\hbar\Gamma$(QW1) = 178 $\mu$eV,$\hbar\Gamma$(QW2) = 217 $\mu$eV, $\hbar\Gamma$(QW3) = 307 $\mu$eV; (d) sample P561: $\hbar\Gamma$ = 152 $\mu$eV. Insets show the narrowest lines magnified.}
\label{spectra}
\end{figure}

A set of reflection spectra was measured in different spots of each sample. The samples with SQW nominally having $L_Z < 4$ nm and indium concentration $x < 0.04$ demonstrate the most narrow exciton lines. Spectra from the four most perfect SQW samples P550, P554, P566, and P561 are presented in Fig. 2(a)-(d) by red solid lines. Typical spectrum displays heavy-hole, light-hole as well as higher energy exciton transitions located around the GaAs bulk exciton resonance ($E>1.514$ eV). The presented spectra demonstrate exciton linewidths narrower than those reported in literature \cite{Poltavtsev1, Zhang, Prineas1, Hubner2}. In particular, the heavy-hole exciton line of the sample P554 has FWHM $\hbar\Gamma=130$ $\mu$eV [Fig.~\ref{spectra}(b)] and for the sample P561 $\hbar\Gamma=152$~$\mu$eV [Fig.~\ref{spectra}(d)]. We note that these linewidths include temperature-dependent part, which can be extracted by measuring the temperature dependence of reflection \cite{Poltavtsev1}. It was experimentally checked that this broadening is $\hbar\Gamma_2\approx13$ $\mu$eV for $T=9$ K and linearly depends on temperature up to 40 K with a coefficient of about 1.5 $\mu$eV/K.

\begin{figure}[t]
\includegraphics[width=\linewidth]{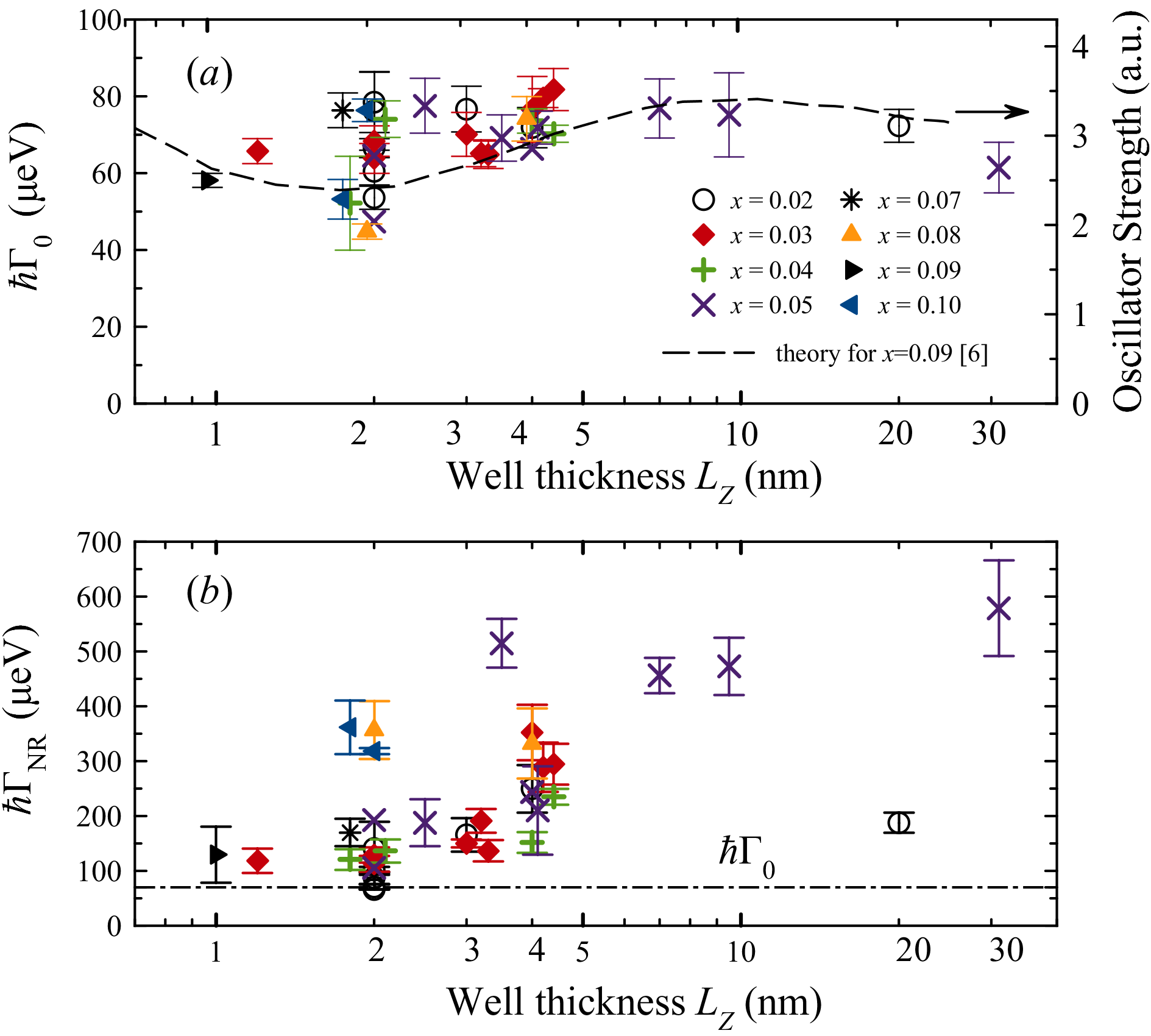}
\caption{(Color online) Experimental dependence of (a) exciton radiative linewidth and (b) non-radiative linewidth on QW thickness and indium concentration obtained on samples listed in Table \ref{tabular:samples_data}.  {\it Dashed} line in (a) is the theoretically calculated oscillator strength taken from \cite{DAndrea}.  {\it Dash-dotted} line in (b) is averaged value of measured $\hbar\Gamma_0$ (70 $\mu$eV).}
\label{gamma_0}
\end{figure}

For all reflection spectra resonant features related to the lowest heavy-hole exciton transition from each SQW were analyzed. For the case of isolated resonant lines, Lorentz approximation was applied to extract values of resonant reflection coefficient $K_{RR}$ and resonance linewidth $\Gamma$; then, Eqs. (\ref{Gamma_zero}) were used to calculate radiative and non-radiative linewidths. The spectra containing a number of resonant lines were simulated with the transform matrix method to obtain the same quantities, $\Gamma_0$ and $\Gamma_{NR}$. The simulated spectra are shown in Fig.~\ref{spectra}(a)-(d) with black dashed lines. The results of experimental measurements of $\Gamma_0$ and $\Gamma_{NR}$ as a function of QW thickness and indium concentration are presented in Fig.~\ref{gamma_0}. They reveal weak dependence of $\Gamma_0$ on both parameters smoothed by the measurement scatter, and can be approximated by a constant value $\hbar\Gamma_0\approx70$ $\mu$eV. Our experimental results are in good qualitative agreement with the theoretical calculations of exciton oscillator strength performed for (In,Ga)As/GaAs SQW \cite{DAndrea}. Although error of each experimental point averaged over $6\div8$ spectra is rather small due to high local homogeneity of studied quantum wells, there is some scatter of points, which we ascribe to a random deviation between nominal and real QW layer thicknesses and indium concentrations.

\begin{figure}[t]
\includegraphics[width=\linewidth]{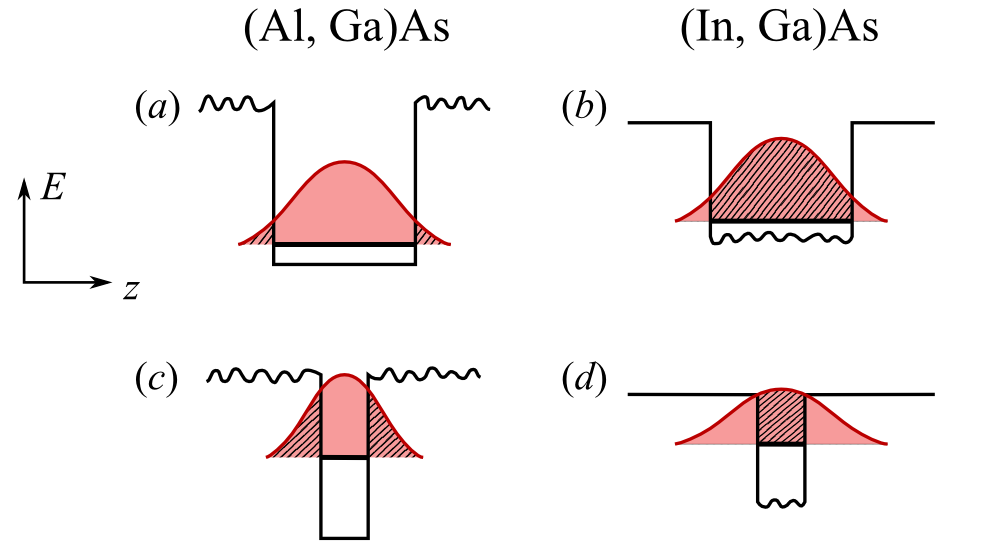}
\caption{Schematic representation of the exciton wave-function penetration into the disordered AlGaAs barriers of thick (a) and thin (c) GaAs QWs and in the GaAs barriers of disordered thick (b) and thin (d) InGaAs QWs. The parts of exciton wave-function interacting with the disorder are marked by a hatch.}
\label{qw_schemes}
\end{figure}

The fact that $\Gamma_0$ is practically universal property of reflection spectra measured on different (In,Ga)As/GaAs QWs results in simplification of Eq. (\ref{KRR}) connecting now the amplitude of exciton resonant line, $K_{RR}$, with the broadening factor $\Gamma_{NR}$. Thus, one may compare both quantities among different QWs. According to our measurements, the lowest $\Gamma_{NR}$ is achieved on quantum wells having 2 nm thickness and indium concentration of about 0.02. On a number of samples, we obtained $\Gamma_{NR}$ comparable to $\Gamma_0$ or even less, as it is seen from Fig.~\ref{gamma_0}(b). According to  Eq. (\ref{KRR}), when $\Gamma_{NR}=\Gamma_0$ theoretical value of the reflection coefficient $K_{RR}$ is 0.25. In the P561 sample, the reflection amplitude reaches value 0.32 and $\Gamma_{NR}=0.79\Gamma_0$. The main result we obtained here is approaching of inhomogeneous broadening to a minimal value with decreasing of (In,Ga)As/GaAs QW thickness, which is opposite to the case of (Al,Ga)As/GaAs QWs, for which thinner QWs possess higher inhomogeneity. This can be understood with the help of the scheme shown in Fig.~\ref{qw_schemes}. In the (Al,Ga)As structures, diminishing of QW thickness causes the larger penetration of the exciton wave-function to the disordered ternary alloy of the barriers [see Fig.~\ref{qw_schemes}(a) and (c)] giving rise to the higher inhomogeneity. In case of (In,Ga)As structures, narrowing of QW layer leads to the escape of exciton wave-function from the disordered InGaAs region to the GaAs barrier, as shown in Fig.~\ref{qw_schemes}(b) and (d), with subsequent decrease of inhomogeneity.

\begin{figure}[t]
\includegraphics[width=\linewidth]{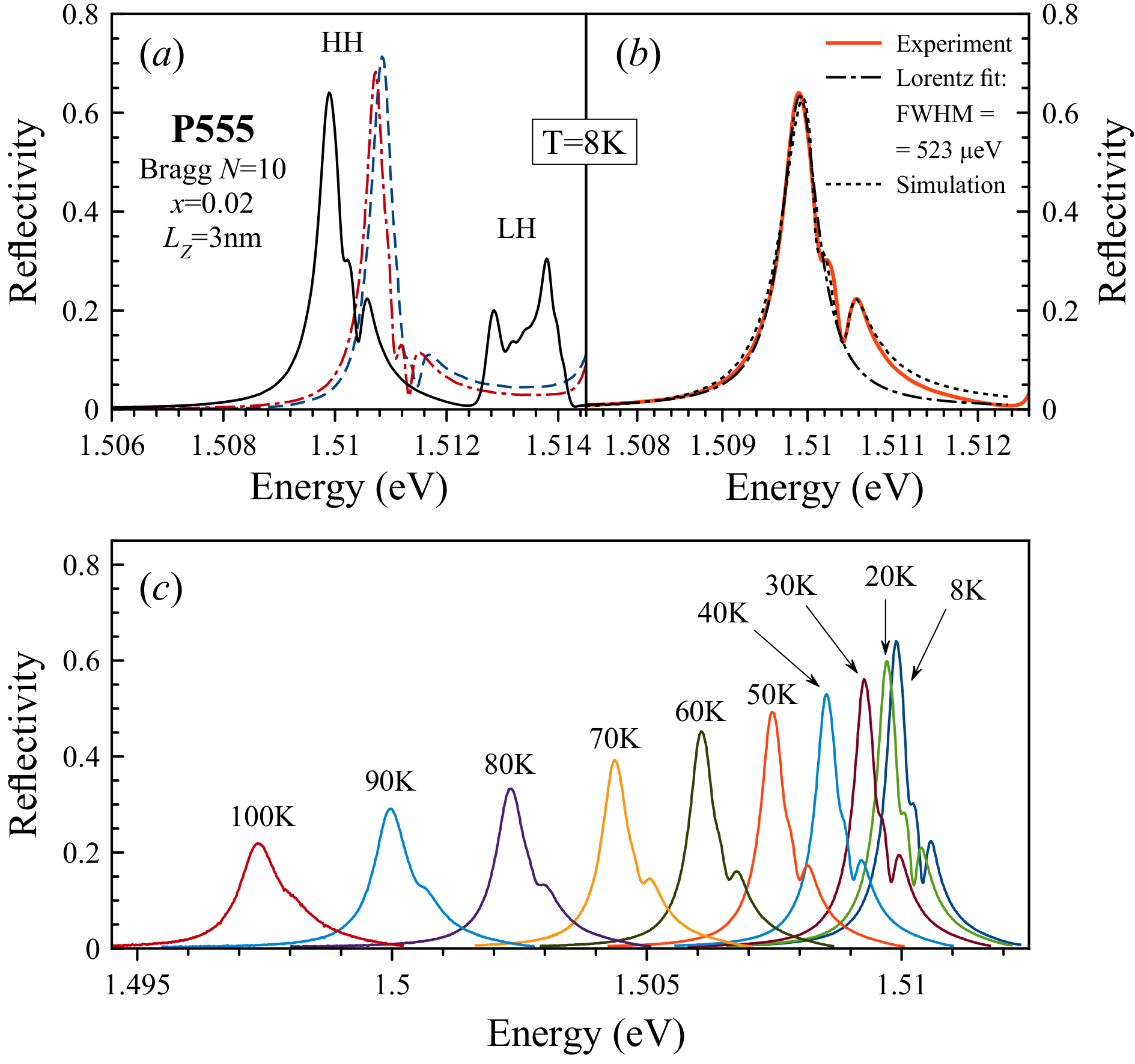}
\caption{(Color online) Reflection spectroscopy of the MQW Bragg structure: (a) the reflection spectra taken at $T=8$K at different spots; (b) simulation of experimental spectrum ({\it red solid}) by fitting with simple Lorentzian ({\it blue dash-dot}) and using the transfer matrix method ({\it black dot}) giving for each QWs $\hbar\Gamma_0=68$ $\mu$eV; (c) the reflection spectrum as a function of temperature.}
\label{mqw_spectrum}
\end{figure}

\subsection{\label{sec:level1}Reflection from MQW Bragg structure}

To get enhanced reflection from quantum well, we used the MQW Bragg structure. The sample P555 containing $N=10$ spatially periodic 3 nm QWs with indium concentration $x=0.02$ has been designed to perform high efficient reflection in the Brewster geometry. Since no substrate rotation during the growth process was used, the sample had spatial gradient in optical properties and, hence, possessed a pronounced inhomogeneity. It was possible, however, to locate a macroscopic area of the sample, where Bragg condition, i.e. separation of QW layers by half-wavelength period, was fulfilled. The spectra presented in Fig.~\ref{mqw_spectrum}(a) were measured at different sample spots and demonstrate resonant reflection coefficient up to 0.7. The spectrum presented with black solid line was analyzed with transfer matrix method described above using minimal number of free parameters. The simulation result is shown in Fig.~\ref{mqw_spectrum}(b) with the black dotted line together with original spectrum (red solid). Deviation of the real spectrum from the Lorentz shape can be explained by two assumptions we employ: 1) there is a small mismatch of MQW spatial period responsible for the slight shape distortion; 2) one of ten QWs is split-off the main resonance, thus, leading to an additional peak at the right tail of resonant line. For all single QWs the value of radiative linewidth was set equal. Also, nine top QWs were supposed to have equal non-radiative linewidths, energy positions and barrier thicknesses. These parameters for the tenth (deepest) QW were left free. Consequent fitting procedure gave value $\hbar\Gamma_0=68$ $\mu$eV for radiative linewidth of each QW, which is in good agreement with the results obtained on a set of SQW structures. Deviation within 2\% is achieved between the fitted spatial MQW period (115 nm) and the Bragg period calculated for the Brewster geometry: $\lambda/2n_0cos\varphi=112.8$ nm, where $\lambda=821$ nm is resonant wavelength, $n_0\approx3.5$ is GaAs refractive index, and $\varphi\approx16^{\circ}$ is light propagating angle inside the structure.

\begin{figure}[h]
\includegraphics[width=\linewidth]{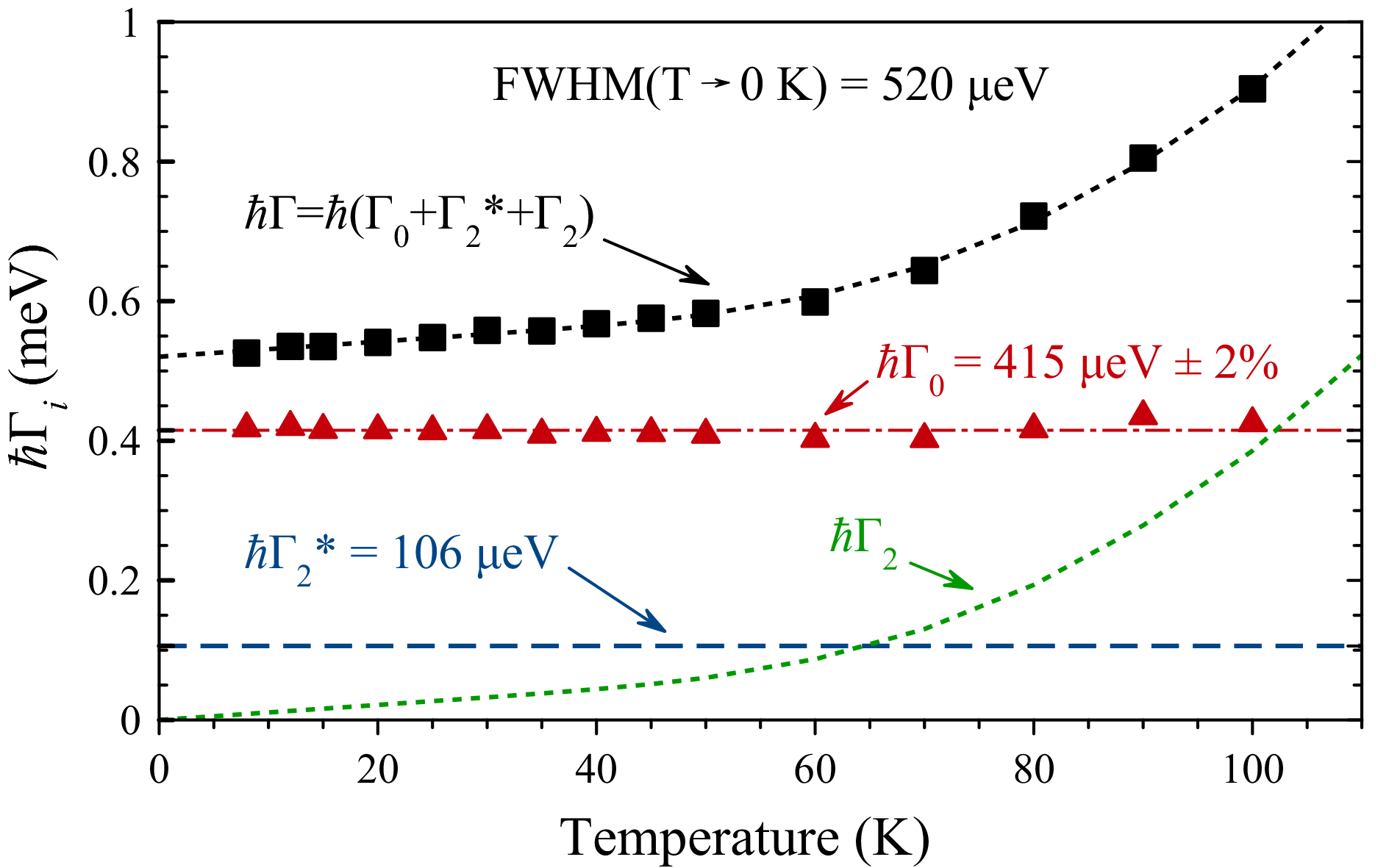}
\caption{(Color online) The experimental temperature dependence of Lorentz FWHM ({\it black squares}), fitting with Eqs. (\ref{Gamma_T}) ({\it black dash}), extracted radiative linewidth, $\hbar\Gamma_0=415$~$\mu$eV $\pm$ 2\% ({\it red triangles with dash-dotted guide}), inhomogeneous broadening, $\hbar\Gamma_2^*=106$ $\mu$eV ({\it blue dash}) and homogeneous temperature-dependent broadening $\hbar\Gamma_2$ ({\it green dash})}
\label{temperature}
\end{figure}

The spectrum under consideration can be satisfactorily approximated by a simple Lorentzian [dash-dotted line in Fig.~\ref{mqw_spectrum}(b)] with FWHM = 523 $\mu$eV characterizing the super-radiant mode linewidth. With this simplified approximation, we may try to perform further analysis of the resonant line structure based on measurements of reflection spectrum as a function of temperature \cite{Poltavtsev1}. This method allows one to extract the radiative linewidth of the super-radiant mode and to separate non-radiative linewidth onto temperature-dependent homogeneous broadening $\Gamma_2(T)$ and inhomogeneous broadening $\Gamma_2^*$. Evolution of the reflection spectrum measured at the same spot with increasing temperature is displayed in Fig.~\ref{mqw_spectrum}(c). With rising temperature, exciton line decreases in amplitude, becomes broader and experiences red shift. Measured FWHM of the exciton line as a function of temperature is displayed in Fig.~\ref{temperature} with black squares. It can be fitted with the following expression considering that only $\Gamma_2$ is  temperature-dependent \cite{Koteles}:

\begin{equation}\label{Gamma_T}
\begin{split}
\Gamma(T)=\Gamma_0+&\Gamma_2^*+\Gamma_2(T),\\
\Gamma_2(T)=\gamma_{ac}T+\Gamma_{LO}/&\left[exp(\hbar\omega_{LO}/k_BT)-1\right],
\end{split}
\end{equation}

\noindent where $\hbar\gamma_{ac}=1.1$ $\mu$eV/K and $\hbar\Gamma_{LO}=11.5$ meV are fitted coupling constants of exciton-acoustic-phonon and the exciton-optical-phonon interactions, $\hbar\omega_{LO}\approx 36$ meV is the optical phonon frequency in GaAs \cite{Chandrasekhar}, and $k_B$ is the Boltzmann constant.

Using Eqs. (\ref{Gamma_zero}) and (\ref{Gamma_T}), the radiative as well as non-radiative linewidths can be extracted: $\hbar\Gamma_0=415$ $\mu$eV, $\hbar\Gamma_2^*=106$ $\mu$eV. These values are shown in Fig.~\ref{temperature} by red triangles and blue dashed line, respectively. The fact that $\Gamma_0$ appeared to be a well-fixed constant serves as a reliable argument that we are dealing with a single oscillator (super-radiant mode) possessing enhanced oscillator strength. The radiative decay rate of this super-radiant mode is about four times higher than the rate of non-radiative damping caused by the inhomogeneity.

To extract the individual QW contribution $\Gamma_0'$ to the total radiative linewidth of the super-radiant mode, we may use the following function describing the line shape \cite{Ivchenko4}:

\begin{equation}\label{R_N}
R(\omega)=\frac{(N\Gamma_0')^2}{(\omega-\omega_0)^2 +(N\Gamma_0'+\Gamma_{NR})^2}
\end{equation}

\noindent Here, $N=9$ is the number of QWs involved in the enhanced reflection. From here we get $\hbar\Gamma_0'\approx46$ $\mu$eV, which is somewhat smaller than the value, calculated with the transfer matrix method (68 $\mu$eV). This difference, as we assume, comes from the mismatch of the real spacing between QWs with calculated Bragg-period resulting in loss of super-radiant mode efficiency.

We can compare our results with the data obtained by Prineas et. al. \cite{Prineas2,Hubner2} on a set of 8.5 nm In$_{0.04}$Ga$_{0.96}$As/GaAs MQW Bragg structures having anti-reflection coating and studied in the normal incident geometry. The authors have demonstrated linear growth of total linewidth for increasing QW number $N$. For $N=10$ the resonant reflection coefficient was about 0.36 and FWHM$\approx$1 meV, which are quite different as compared to the values obtained on our sample (0.7 and 0.52 meV, respectively). This difference, as we believe, comes from the relatively strong inhomogeneous broadening present in the spectra shown in \cite{Prineas2,Hubner2}. We assume that this inhomogeneity is conditioned by the larger thickness of QWs (8.5 nm) and higher indium concentration (0.04) as compared to 3 nm and 0.02, respectively, in our case.

\section{\label{sec:level1}Conclusion}

In conclusion, we have performed reflection spectroscopy from shallow InGaAs QWs and MQW Bragg structure in the Brewster angle geometry. Background-free spectra of QWs with low concentration of indium demonstrate heavy-hole exciton resonances with linewidth below 0.2 meV and resonant reflection coefficient up to 0.32 proving equal roles of the radiative and non-radiative mechanisms of exciton decay. We have shown that the dependence of oscillator strength on SQW thickness and indium concentration within 0.02-0.10 is weak. On high-quality MQW Bragg structure we have studied the enhancement of radiative contribution to the resonance linewidth. We have checked that the resultant super-radiant mode behaves as a single oscillator conserving its oscillator strength at temperature ranging between 8 and 100 K. Both considered possibilities of increasing the role of the radiative contribution to the exciton linewidth are important for application of QW-based materials in optical processing and storage devices, for which efficiency of the coherence transfer in the process of mutual exciton-photon interconversion, $F$, as well as high non-linearity are the key properties \cite{Oves1,Oves2}. QWs featuring narrow exciton lines are also appealing for fast switching applications and devices requiring a dense spectral lines design.

\section*{\label{sec:level1}Acknowledgments}

Authors thank Prof. Dmitri Yakovlev and Gleb Kozlov for instructive discussions. The financial support from the Russian Ministry of Education and Science (Contract No. 11.G34.31.0067 with SPbU and leading scientist A. V. Kavokin) is acknowledged. The authors also acknowledge SPbU for research Grants No. 11.38.67.2012 and No. 11.38.213.2014. This work has been partially funded by Skolkovo Institute of Science and Technology (Skoltech) in the framework of the SkolTech/MIT Initiative and by the EU FET-program SPANGL4Q. S.V.P. thanks RFBR for the partial financial support of this work (Contract No. 14-02-31735 mol-a). It was carried out using the equipment of SPbU Resource Center ''Nanophotonics'' (photon.spbu.ru).

\end{document}